\documentclass[10pt,letterpaper]{article}
\usepackage{opex3}
\usepackage{float}
\usepackage{epsfig}

\begin{document}


\title{\emph{In-situ} determination of astro-comb calibrator lines to better
  than 10~cm~s${}^{-1}$}

\author{Chih-Hao Li,${}^{1,2}$
Alexander G. Glenday,${}^{1,2}$
Andrew J. Benedick,${}^{3}$
Guoqing Chang,${}^{3}$
Li-Jin Chen,${}^{3}$
Claire Cramer,${}^{1}$
Peter Fendel,${}^{3,4}$
Gabor Furesz,${}^{2}$
Franz X. K\"artner,${}^{3}$
Sylvain Korzennik,${}^{2}$
David F.\ Phillips,${}^{2}$
Dimitar Sasselov,${}^{2}$
Andrew Szentgyorgyi,${}^{2}$ and
Ronald L.\ Walsworth${}^{1,2}$
}


\address{${}^1$Department of Physics,
Harvard University}
\address{${}^2$Harvard-Smithsonian Center for
Astrophysics}
\address{${}^3$Massachusetts Institute of Technology}
\address{${}^4$Menlosystems, Incorporated}

\date{\today}
\begin{abstract}
  Improved wavelength calibrators for high-resolution astrophysical spectrographs will be essential for
  precision radial velocity (RV) detection of Earth-like exoplanets and direct
  observation of cosmological deceleration.  The astro-comb is a combination of an octave-spanning
  femtosecond laser frequency comb and a Fabry-P\'erot cavity used to achieve
  calibrator line spacings that can be resolved by an astrophysical
  spectrograph. Systematic spectral shifts associated with the cavity can be
  0.1-1~MHz, corresponding to RV errors of 10-100~cm/s, due to the dispersive properties of the cavity
  mirrors over broad spectral widths.  Although these systematic shifts are very stable, their
  correction is crucial to high accuracy astrophysical
  spectroscopy.  Here, we demonstrate an \emph{in-situ} technique to
  determine the systematic shifts of astro-comb lines due to finite Fabry-P\'erot cavity dispersion. The technique is practical for implementation at a telescope-based spectrograph to enable wavelength calibration accuracy better than 10 cm/s.
\end{abstract}

\vspace{2ex}
\bibliographystyle{osajnl}



\section{Introduction}

High precision wavelength calibrators for astrophysical spectrographs
will be key components of new precision radial velocity (RV)
observations, including the search for Earth-like extra-solar planets
(exoplanets)~\cite{Lovis2006} and direct observation of cosmological
acceleration~\cite{Sandage1962, Loeb1998}. Recent work has
demonstrated the potential of octave-spanning femtosecond laser
frequency combs~\cite{Udem2002} (``astro-combs") to serve as
wavelength calibrators for astrophysical spectrographs providing RV
sensitivity down to 1~cm/s~\cite{Murphy2007, Schmidt2007, Araujo2007,
  Osterman2007, Li2008, Steinmetz2008, Braje2008}.  Exoplanet searches
place stringent demands upon such calibrators.  For example, the RV
amplitude of the reflex motion of a solar-mass star induced by an
Earth-mass planet in an orbit within the habitable zone is about
10~cm/s. The current state of the art astrophysical spectrograph,
HARPS, has demonstrated stellar RV sensitivity
$\approx60$~cm/s~\cite{Lovis2006}, largely limited by its thorium
argon lamp calibrator~\cite{Lovis2006b, Udry2007}. These calibrators
are limited by their unevenness in line spacing and intensity as well
as the slow variation of their line wavelengths with time.
An astro-comb provides emission lines with uniform intensity and
controllable spacing, which can be referenced to atomic frequency
standards and the global positioning system (GPS), yielding excellent
long-term stability and reproducibility.

\begin{figure}
\centering
\includegraphics[width=5 in]{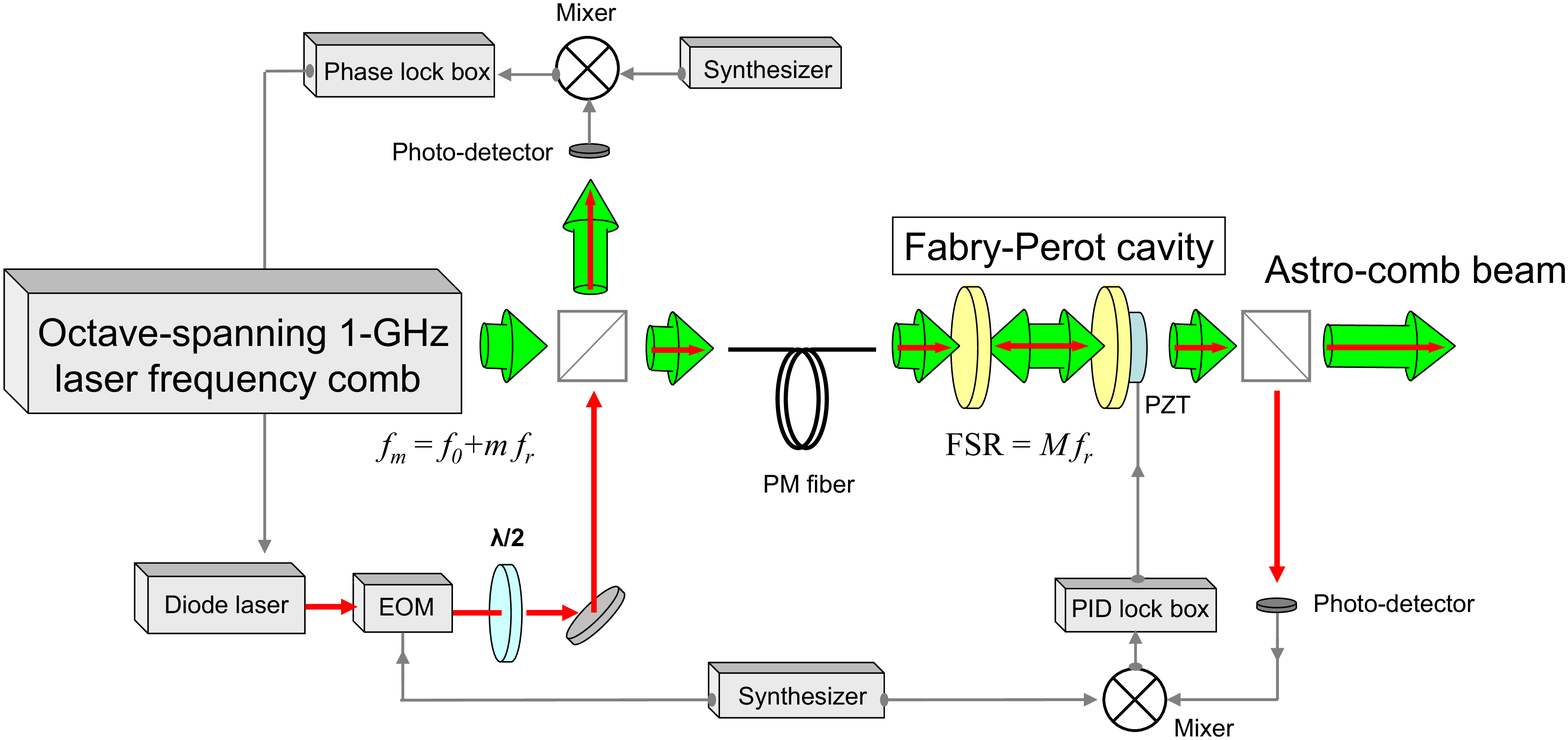}
\caption{Block diagram of astro-comb used in measurements reported here. A 30-GHz line spacing is
  generated from a 1-GHz source comb mode-filtered by a Fabry-P\'erot
  Cavity stabilized to an injected diode laser via the
  Pound-Drever-Hall method in transmission. The diode laser is phase
  locked to one source comb line, while the source comb is stabilized
  to an atomic frequency reference.} \label{fig:astro-comb setup}
\end{figure}

To date, astro-combs consist of a combination of an octave-spanning
femtosecond laser frequency comb (source comb) and a Fabry-P\'erot
cavity (FPC), see Fig.~\ref{fig:astro-comb setup}.
Spectral lines generated by the source comb are spaced by the pulse
repetition rate ($f_r$), currently $\approx 1$~GHz, which results in a line spacing too dense to be resolved by broadband astrophysical spectrographs~\cite{Steinmetz2008}. The FPC serves as
a mode filter with a free spectral range (FSR) set to an integer multiple of the
repetition rate, FSR=$M f_r$, with $M \approx 10-100$, depending on the
spectrograph resolution.
Ideally, the FPC passes only every $M^{th}$ source comb spectral line, providing thousands of calibration lines well matched to a practical spectrograph's resolution, with fractional frequency uncertainty limited only by the stability of the RF reference used to stabilize the source comb and FPC. This frequency uncertainty can be $<10^{-11}$ using commonly available atomic clock technology, which corresponds to
$\sim3$ kHz uncertainty in the optical frequency or 0.3 cm/s in RV precision. However, because the spectrograph fails to resolve neighboring source comb
lines, finite suppression of these neighboring lines by the FPC affects the
lineshape and potentially the centroid of measured astro-comb
lines.
For example, in the results presented here, source comb modes
neighboring the astro-comb line, with intensities after passing through
the FPC that differ by 0.1\% of the main astro-comb peak, shift the
measured line centroid by 1 MHz, which corresponds to an RV systematic error of 1 m/s.
In practice, such systematic RV shifts are
inevitable over spectral bandwidths of 1000~\AA\ due to the
dispersive properties of the mirrors of the FPC. Although these systematic shifts can be very stable over timescales of years, the correction of such shifts is crucial to high accuracy astrophysical spectroscopy.

In this paper, we demonstrate an \emph{in-situ} technique to determine the systematic shifts of astro-comb lines due to FPC dispersion, which can be applied at a telescope-based spectrograph to enable wavelength calibration accuracy better than 10 cm/s. By measuring
the intensity of astro-comb lines as the FPC length is adjusted, we determine
(i) the offset of each FPC resonance from the nearest source comb
line; (ii) FPC finesse as a function of wavelength; and
(iii) the intensity of the astro-comb lines and their neighboring (normally suppressed) source comb lines. These parameters
can be determined quickly and reliably over the full 1000~\AA\ bandwidth of the astro-comb with only $\approx 50$ measurements at slightly different FPC lengths,
and can be performed quickly ($<1$ hour) and reliably.  The measurement has also
been performed with a lower resolution commercial optical spectrum
analyzer with consistent results.
The astro-comb line characterization technique presented here builds on past work in which femtosecond lasers coupled to swept cavities were used to study both the medium in which the cavity was immersed and the cavity mirrors ~\cite{Ye05,Hansch06}.

\section{Model}
\label{s.model}
\begin{figure}
\centering
\begin{tabular}{c}
\epsfig{file=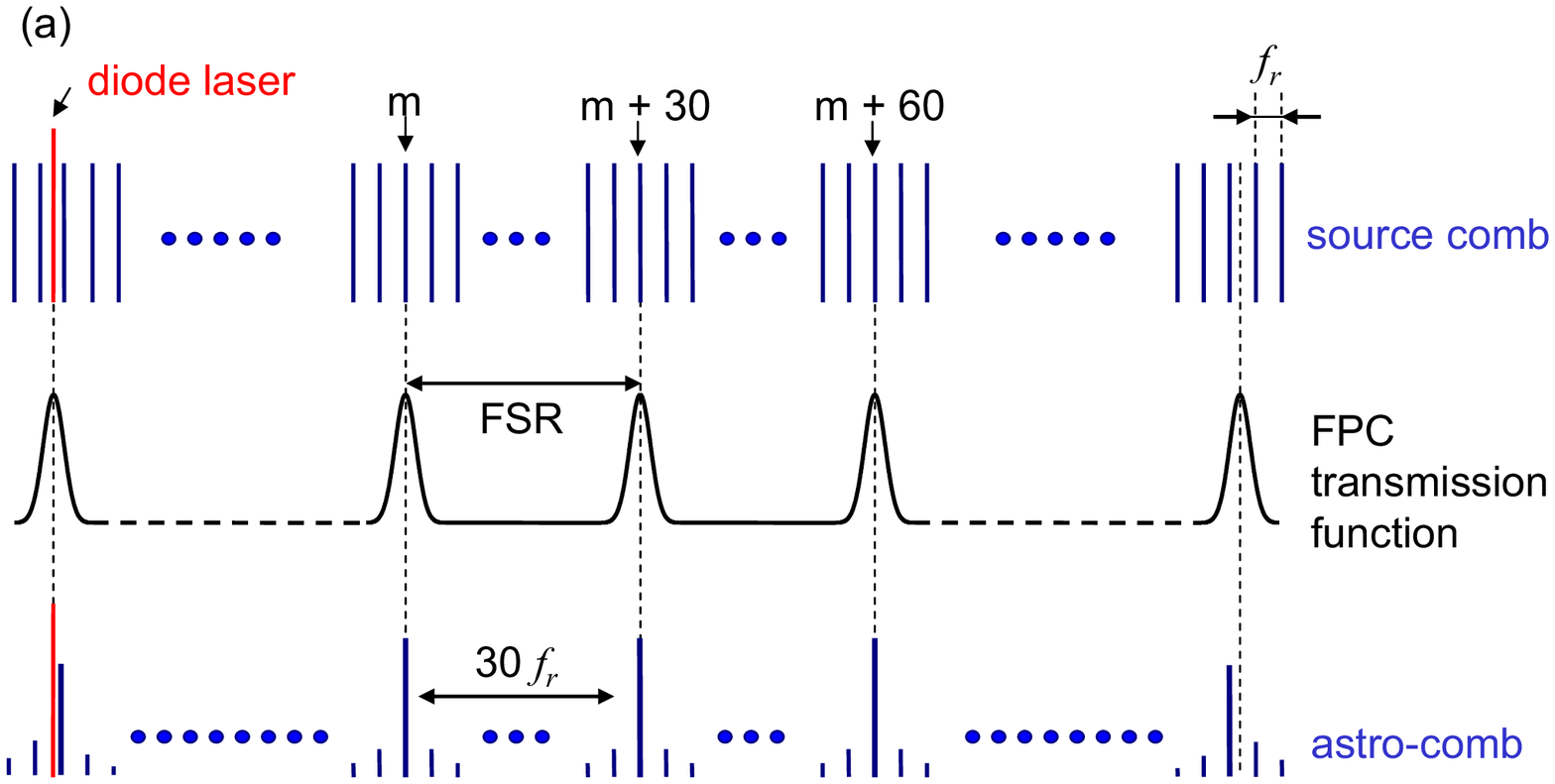, width = 4.5 in}\\
\epsfig{file=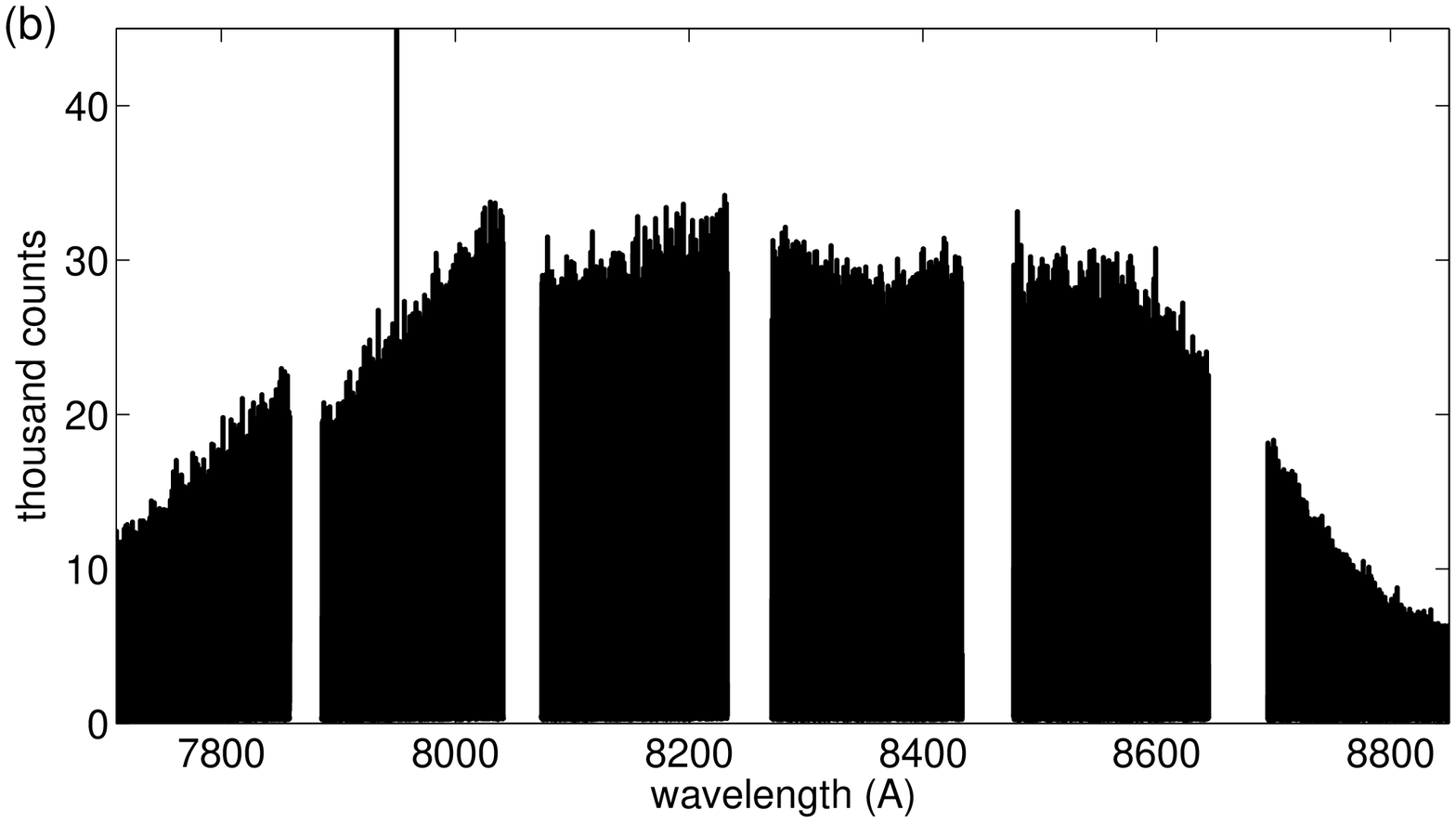, width = 4.5 in}\end{tabular}
\caption{(a) Sketch of (i) source comb lines; (ii) Fabry-P\'erot
  Cavity (FPC) resonances with a Free Spectral Range (FSR) 30 times
  the repetition rate; and (iii) astro-comb spectrum. For these
  settings, one source comb line out of 30 passes the FPC
  unattenuated, producing an astro-comb spacing of $\sim30$~GHz that
  can be resolved by a high-resolution astrophysical spectrograph. The
  wavelength of the reference diode laser is shown slightly detuned
  from the source comb to compensate for FPC dispersion between the
  diode laser wavelength and the central astro-comb wavelength as
  described in the text. (b) Astro-comb lines measured with the TRES
  spectrograph cover a bandwidth $\approx 1000$ \AA. The gaps in the
  spectrum are not measured by the spectrograph due to light in these
  spectral regions not focusing onto its CCD.} \label{fig:comb_FP}
\end{figure}

Imperfect suppression of unwanted source comb lines, e.g., due to finite FPC finesse, affects the astro-comb lineshape
observed on a spectrograph. The lineshape can be modeled with
knowledge of the FPC properties including mirror reflectivity and
round trip phase delay.  The intensity of a source comb line after the FPC is
\begin{equation}
I'_m = I_{m}
\frac{ T_m}{1+ (2 F_m / \pi)^2 sin^2(\phi_m/2)},
\label{eqn:FP_transmission}
\end{equation}
where $I_{m}$ is the intensity of the source comb line of optical
frequency $f_m$; $T_m$ is the resonant transmission of the FPC at
optical frequency $f_m$; $F_m$ is the finesse of the FPC near
frequency $f_m$; and $\phi_m$ is the round trip phase delay. The phase delay may be expressed as~\cite{YarivPhotonics}
\begin{equation}
\phi_m = 2\pi\left(\frac{2L}{c} f_m n_m + \int^{f_m}_{0} \tau(f)
  \, df\right), \label{eqn:phi_0_f}
\end{equation}
where $L$ is the length of the cavity, $c$ is the speed of light in
vacuum, $n_m$ is the refractive index of the medium inside the cavity (air, vacuum, etc.) at optical frequency
$f_m$, and $\tau(f)$ is the frequency-dependent group delay of the
mirrors. The first term in parentheses in Eq.~(\ref{eqn:phi_0_f}) is the distance between the mirrors expressed in wavelengths; while the second term, the integral of $\tau(f)$, is the phase delay of the mirrors and
represents the frequency-dependent penetration distance of light into the mirror.  Maximum transmission occurs when a source comb line is resonant with the FPC, or equivalently $\phi = 2 \pi q$, with $q$ an integer (see Fig.~\ref{fig:comb_FP}a). The FSR, $\Delta$, is the frequency difference between
two consecutive FPC resonant frequencies.
Assuming that $\tau$ varies slowly with frequency, the FSR can be approximated by
\begin{equation}
\Delta_m \approx \left[\frac{2L}{c} \left( n_m + f_m \frac{dn_m}{df}\right) + \tau_m \right]^{-1}.
\end{equation}

For the astro-comb shown in Fig.~\ref{fig:astro-comb setup}, which we
deployed as a wavelength calibrator for the Tillinghast Reflector
Echelle astrophysical Spectrograph (TRES)~\cite{Furez2009}, typical
operational parameters are $\Delta_m \approx 31$ GHz and $F_m \approx
180$, with source comb lines nearest to the astro-comb peak suppressed
by $\approx 22$ dB (see Fig.~\ref{fig:comb_FP}b and
Section~\ref{Section: Experiment}). This imperfect suppression
leads to systematic inaccuracies in astro-comb line centers at the 50
cm/s level, as observed on the TRES spectrograph across a 1000 \AA\
bandwidth.
The effects of these systematic shifts can be characterized by
determining the FPC finesse ($F_m$) and the frequency difference
between astro-comb lines and the FPC resonance over the full spectral
width.
As described below, compensation can then be applied to determine wavelength
solutions (conversions from spectrograph pixel number to wavelength)
with accuracy at least an order of magnitude below the 50 cm/s level
of these systematic shifts.
%

We determine $F_m$ and the frequency offset of the FPC
resonance from the astro-comb lines by adjusting the length of the
FPC and measuring the intensities of astro-comb lines over the full spectral bandwidth. When we change the frequency of one FPC resonance from $f_d$ to
$f_d + \delta \! f$ the cavity length changes by
\begin{eqnarray}
  \delta L & \approx & - \frac{c}{2n_df_d}\left[ \tau_d +
    \frac{2L}{c}\left( n_d + f_d \frac{dn}{df}|_{f_d} \right)
  \right]\delta\! f \\
  &\approx& - \frac{c}{2n_df_d\Delta_d}\delta \! f,
\end{eqnarray}
where $\Delta_d$ and $\tau_d$ are the FSR of the FPC and the phase
delay of the mirrors, respectively, at the frequency of the controlled
resonance.  The change of the cavity length leads to a change of the
round trip phase delay of an astro-comb line $f_m$ by
\begin{equation}
  \delta\phi_m = \frac{4 \pi n_m f_m}{c} \delta L \approx -2\pi
  \frac{n_m f_m}{n_d f_d \Delta_d} \delta \! f.
\end{equation}
As a result, the intensity of the astro-comb line varies with the
change of the frequency of one FPC mode according to
Eq.~(\ref{eqn:FP_transmission}), from which we can derive the finesse and the phase errors at all astro-comb line frequencies.

\section{Experiment} \label{Section: Experiment}

\begin{figure}
\centering
\includegraphics[width=4.5in]{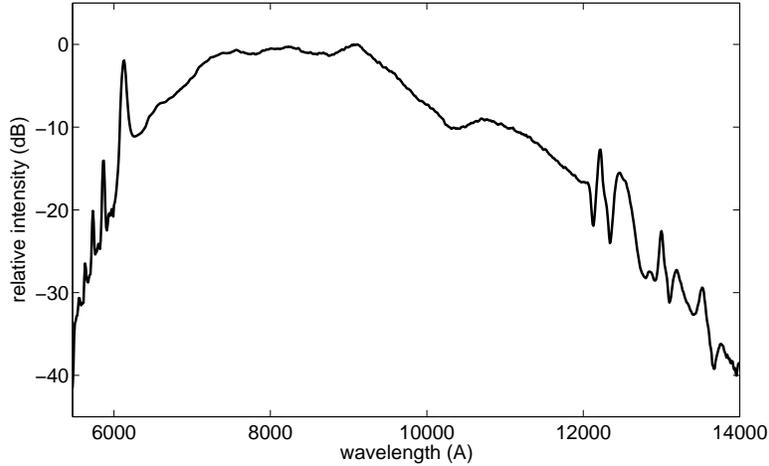}
\caption{Source comb spectrum as measured with a commercial optical
  spectrum analyzer (ANDO AQ6315, R$\approx1,000$). The source comb consists of $\sim10^5$ narrow
  lines (width $<1$ MHz), equally spaced in frequency ($\approx
  1$~GHz, not resolved here), and spanning more than an octave between
  6000~\AA\ and 12000~\AA.}
\label{fig:source comb spectrum}
\end{figure}

The astro-comb employed in our experimental demonstrations is shown schematically in Fig.~\ref{fig:astro-comb
  setup}. The octave-spanning source comb spectrum (Fig.~\ref{fig:source comb spectrum}) is generated by a mode-locked titanium-sapphire femtosecond laser (Octavius, Menlosystems, Inc.) with repetition rate $f_r\approx 1$~GHz.  The absolute
frequencies of the comb lines can be expressed as $f_m=f_0+m~f_r$,
where $m$ is an integer and $f_0$ is the carrier-envelope offset
frequency. A PIN diode detects $f_r$, which is then stabilized by
adjusting the laser cavity length.  The $f$-$2f$ self-referencing
method is used to produce a signal at $f_0$ on an avalanche photodiode: comb lines
around 11400 \AA\ are frequency doubled in a 1 mm thick LBO (Lithium Triborate) crystal
and beat with comb lines around 5700 \AA. $f_0$ is then stabilized by
intensity modulation of the 7.6 W, 5320 \AA\ pump laser. Both $f_r$
and $f_0$ are referenced to low-noise radio-frequency synthesizers,
which are stabilized to a commercial rubidium frequency reference. For our source comb, typical values of these frequencies are: $f_r =
1.042048458$~GHz and $f_0 = 0.09$~GHz, and the resulting linewidth of individual source-comb spectral lines is $<1$~MHz.

The source comb beam passes through the FPC, filtering out unwanted
comb lines and increasing the transmitted line spacing.  Two flat
mirrors with $\approx 98.5$\% reflectivity and minimal group delay
dispersion ($<10$~fs$^2$ from 7500 \AA\ to 9000 \AA) comprise the
FPC. Due to the low dispersion mirrors, the FSR of the FPC is almost
constant within the 1000 \AA\ bandwidth: typically one in thirty comb
lines is resonant with the FPC (Fig.~\ref{fig:comb_FP}a). The measured
finesse of the FPC is $\approx 180$, which is consistent with the
theoretical limit estimated from the mirror reflectivity.

We stabilize the FPC by locking one transmission resonance to an
injected diode laser.
The diode laser is modulated at 30 MHz with an electro-optic modulator
(EOM) and injected into the FPC with a polarization orthogonal to the
comb light.  After the FPC, the diode beam is separated from the comb,
demodulated, and used to derive a feedback signal that stabilizes the
FPC to the diode laser.  The diode laser itself is offset-locked to
one of the source comb lines at a wavelength $\approx7950$~\AA.  The
frequency of the offset lock (see Fig.~\ref{fig:comb_FP}a) is adjusted
to optimize the astro-comb bandwidth, and compensates for FPC
dispersion between the diode laser wavelength and the central
astro-comb wavelength due to mirror properties and air in the cavity
as discussed in Secs.~\ref{s.model} and~\ref{s.results}.
The diode laser wavelength is close to the Rb D1
transition. Absorption spectroscopy of the diode laser light using a
thermal rubidium vapor cell thus identifies the diode frequency to
within one source comb line. The offset lock then determines the
absolute frequency of the diode laser with the accuracy of the
underlying atomic frequency reference.

In its current incarnation as a wavelength calibrator for an
astrophysical spectrograph, the astro-comb spectrum is analyzed by the
Tillinghast Reflector Echelle Spectrograph (TRES)~\cite{Furez2009}, a
fiber-fed multi-order echelle spectrograph at the 1.5 m telescope of
the Whipple Observatory at Mt. Hopkins in Arizona. Spectral dispersion
is provided by an echelle grating with cross-dispersion from a prism
operated in double-pass mode.  TRES covers a spectral bandwidth from
3700~\AA\ to 9100~\AA\ with a resolving power of $R\equiv \lambda /
\delta\lambda \approx 50000$~\cite{Furez2009} corresponding to a
resolution of $\approx7$~GHz at 8000 \AA.  Astro-comb light is
injected into an integrating sphere and then sent to TRES on a
100-$\mu$m multimode fiber. The integrating sphere reduces lineshape
fluctuations due to input-dependent illumination of the spectrograph
optics. Each astro-comb spectral measurement by TRES is typically
integrated for 60~s and recorded on a cooled (100~K) two dimensional
CCD (see Fig.~\ref{fig:comb_FP}b).  Flat-field correction is applied
to the astro-comb spectrum in order to remove artifacts caused by
variations in the pixel-to-pixel sensitivity of the detector and by
distortions in the optical path. The implementation of both the
integrating sphere and the flat-field correction are essential for
high accuracy astrophysical spectroscopy but is not crucial to the
work described in this paper. The separation of the FPC mirrors is set such that astro-comb lines are separated by $\approx4$ resolution elements of the spectrograph. In the measurement of the intensity of each astro-comb lines, typically the counts from 18 CCD pixels around the central pixel are binned from the one dimensional extracted spectrograph spectrum to obtain each measurement points in Fig.~4. The 18 pixel integrations capture all the intensity from each astro-comb line (spectrograph resolution $\approx 6$ pixels) while avoiding cross-talk between neighboring lines (line separation $\approx 24$ pixels).
%
%
We have also varied the binning and found no appreciable change in the result. The length of the FPC was swept (by changing the diode frequency) back and forth through the FPC resonance: we took steps of 80 MHz in one direction and then steps in the reverse direction at intermediate positions. Different step sizes or schemes have led to consistent results.

We also characterized the astro-comb spectrum with a commercial
optical spectrum analyzer (OSA, ANDO AQ6315), which has much lower
resolution ($\approx 300$~GHz) when operated in broadband (1000~\AA)
mode.  There are then $\approx10$ astro-comb lines in one resolution
element.
Extraction of astro-comb intensities from the OSA is significantly simpler than the two dimensional TRES spectrograph, and the OSA, therefore, provides a useful cross-check. We find that all parameters measured with the OSA are consistent with those measured with TRES. Additionally, the calibration procedure has been performed more than 10 times with both the TRES spectrograph and the OSA over 10 days. All measurements are found to be consistent over this time period.

\section{Results and Analysis}
\label{s.results}
\begin{figure}
\centering
  \includegraphics[width=4.5 in]{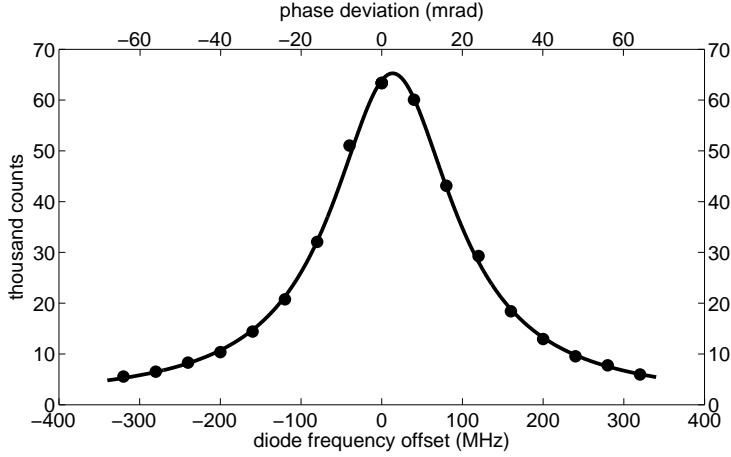}
  \caption{Measurements using the TRES spectrograph of the peak intensity of one astro-comb spectral line as the diode
    laser frequency, and thus the Fabry-Per\'ot cavity length, is varied (solid circles)
    and a fit of Eq.~\ref{eqn:FP_transmission} to the data (solid
    line). Uncertainties in the linewidth and in the offset of the
    peak from zero diode detuning are approximately 1 MHz.
  } \label{fig:data_amp_vs_phase}
\end{figure}

We find good agreement between TRES measurements of the astro-comb spectrum and the model presented above. For example, Figure~\ref{fig:data_amp_vs_phase} shows the measured
variation of the peak intensity of one astro-comb line vs diode laser frequency (and thus cavity length), as well as a fit of
Eq.~\ref{eqn:FP_transmission} to the data. From the fit, we derive the astro-comb resonant intensity $T_m I_{m}$; the FPC finesse $F_m$; and the phase deviation $\delta\phi_m$. The phase deviation is the offset of the round trip phase from an integer multiple of $2\pi$ at an astro-comb line, and is determined from the detuning of the astro-comb peak at frequency $f_m$ from center of the FPC resonance at nominal diode laser setting $f_{\textrm{FPC}}$.  The phase deviation is then given by
\begin{equation}
\delta\phi_m = 2 \pi \ \frac{f_{\textrm{FPC}}-f_m}{\Delta_m},
\end{equation}
where $\Delta_m$ is the free spectral range near frequency $f_m$.  The finesse, $F$, of the FPC can be derived from the fit as
\begin{equation}
F=\frac{2\pi}{\Delta\phi_{\, \textrm{FWHM}}},
\end{equation}
where $\Delta\phi_{\, \textrm{FWHM}}$ is the full width at half
maximum in phase of an FPC resonance.  The uncertainties in measured
FPC resonance centers and widths are approximately 1 MHz,
corresponding to phase uncertainties of 0.2~mrad.  The variation of
these parameters as a function of wavelength provides the information needed to characterize the astro-comb spectrum as a wavelength calibrator.
%

\begin{figure}
\centering
\includegraphics[width=4.5 in]{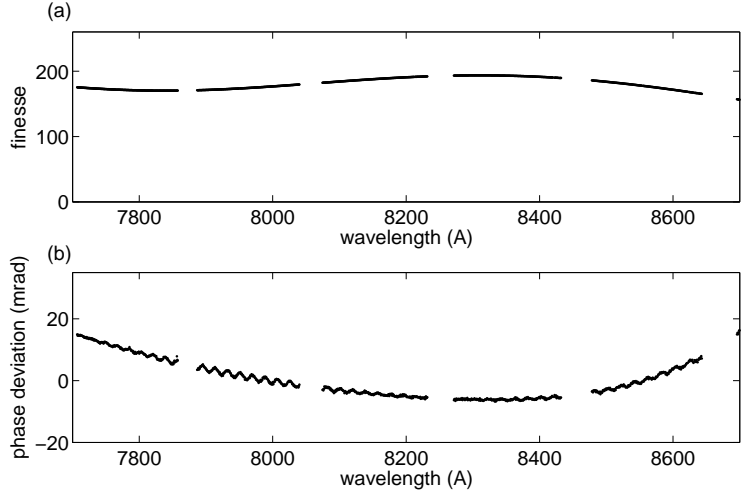}
\caption{ (a) TRES measurements of the finesse of the Fabry-P\'erot cavity and (b)
  phase variation of the astro-comb lines in the range of 7700~\AA\ -
  8700~\AA.  The small, rapid phase variation in (b) is due to 0.1\%
  reflections from the back surfaces of the cavity's antireflection coated mirror substrates (\emph{see text}).  } \label{fig:fitted_phase_error}
\end{figure}

\begin{figure}
\centering
  \includegraphics[width=4.5 in]{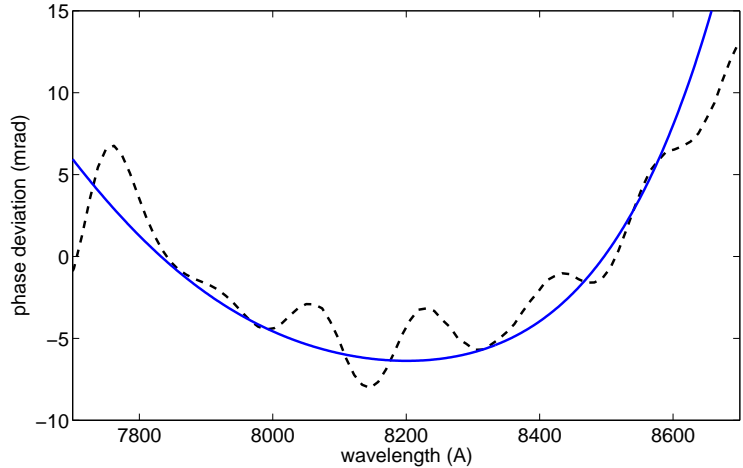}
  \caption{ Polynomial fit of the TRES measurement of the cavity phase deviation as described in the text
    (solid curve) compared to a model derived from air dispersion and
    mirror phase delay as measured with a white light interferometer
    (dashed curve). The constant and linear phase variation which
    depend upon the absolute cavity position and length are removed
    from both curves for clarity.  }
  \label{fig:GDD}
\end{figure}

As shown in Fig.~\ref{fig:fitted_phase_error}, the FPC finesse and phase deviation are found to vary slowly across the entire astro-comb
spectrum, as measured with TRES. The FPC finesse is approximately 180
(Fig.~\ref{fig:fitted_phase_error}a) and the phase deviation of the
FPC relative to astro-comb lines is $<20$~mrad
(Fig.~\ref{fig:fitted_phase_error}b).  Reflections from the back
surfaces of the FPC mirrors used in these measurements
produce the small, rapid phase variations observed in the
figure.
Despite antireflection-coatings the mirror substrates reflected 0.1\% of the incident power.
A simple model with realistic parameters reproduces the observed phase variations and allows systematic effects associated with these
reflections to be eliminated. In future work, slight wedging (for example, 0.5$^{\circ}$) of the
substrates will eliminate these variations.  After fitting the phase
deviation with a sixth order polynomial, we compared our result
with the phase deviation derived from a model of air
dispersion~\cite{Ciddor1996} and the mirror phase delay measured with a white light interferometer (Fig.~\ref{fig:GDD}). The two different
methods agree well, given systematic limitations to the white light interferometer measurement and resultant effect on the model fit.

\begin{figure}
\centering
\includegraphics[width=4.5 in]{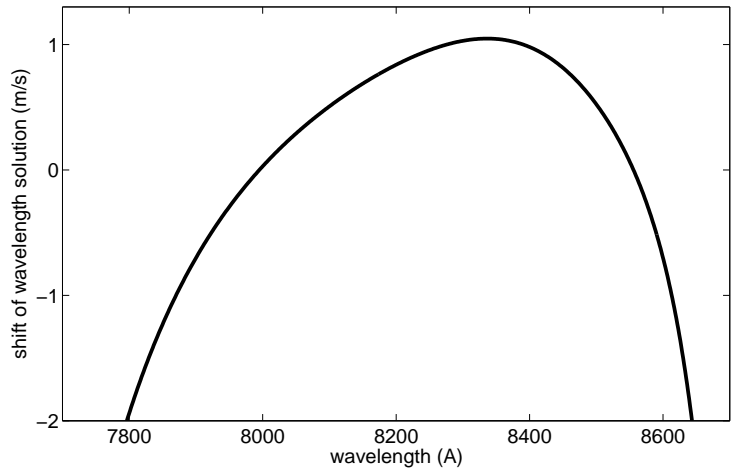}
\caption{Estimate of systematic wavelength (frequency) shift of astro-comb line centers that would be measured at the TRES spectrograph if nearest suppressed source comb lines are not included in the model used to fit for the wavelength solution. } \label{fig:correction}
\end{figure}

Using these \emph{in-situ} measurements of the FPC finesse and the
phase deviation, along with the source comb intensity variation and
resonant FPC transmission, we can determine the suppression of unwanted source comb lines and the resultant frequency shifts of astro-comb line centers.
The variation with optical frequency of source comb line intensity and resonant FPC transmission ($T_m\cdot I_m$ in
Eq.~\ref{eqn:FP_transmission}) is measured by offsetting the diode
laser frequency from its nominal value by the source comb repetition
rate, and thereby tuning the FPC to resonance with a neighboring (and normally suppressed) source comb mode.  The maximal transmission of each transmitted source comb line and thus $T_m\cdot I_m$ is measured; and then the procedure is repeated for all normally suppressed source comb modes.  We find that the variation of $T_m\cdot I_m$ is within our measurement uncertainties ($<1\%$).
Evaluating Eq.~\ref{eqn:FP_transmission} for a single astro-comb line and all of its neighboring (suppressed) source comb lines within one spectrograph resolution element, and determining the net transmitted spectrum weighted by the frequency offset from the central
astro-comb frequency, allows us to evaluate the effective line center shift in the astro-comb spectrum. For moderate finesse and $f_r\gg\Delta_m/F$, the systematic shift $\delta f_{m}$ of an astro-comb line of frequency $f_m$, as measured on a spectrograph, can be approximated
as
\begin{equation}
  \delta f_{m} \approx f_r \left(\frac{\Delta_m/f_r}{2 F} \right)^2 \left[
    \frac{T_{m+1} \, I_{m+1} - T_{m-1}\, I_{m-1}}{T_m \, I_{m}} -
    2\frac{\delta \phi_m}{\pi}   \frac{\Delta_m}{f_r}  \right]
\label{eqn:simpleshift}
\end{equation}
where the term in parentheses parameterizes the mean suppression of
nearest source comb side modes, the first term in square brackets results from
the difference in source comb intensity between the upper and lower
side modes, and the second term in square brackets results
from FPC phase deviation leading to asymmetry of the comb lines
relative to the FPC mode.
The frequency shift in an astro-comb line centroid due to variations
of neighboring source comb line intensities at the 1\% level is thus
$\approx70$~kHz or 7~cm/s. Note that we do not expect neighboring
source comb lines to differ as much as 1\%; also a Gaussian
distribution of comb-line intensities will average away much of this
astro-comb centroid frequency shift. Nonetheless, we expect that a
double pass configuration through the same FPC~\cite{Kirchner2009,
  Steinmetz2009}, moderately higher FPC finesse ($F\approx500$), a
higher repetition rate for the source comb ($f_r\approx 5$~GHz), or
more precise measurements of line to line intensity variations will be
required to assure 1~cm/s accuracy In practical astro-comb
calibrators. In Fig.~\ref{fig:correction}, we show the shift of the
estimated systematic error in the wavelength calibration of TRES as a
function of wavelength, when using our current astro-comb as the
calibration reference, if the effect of nearest suppressed source-comb
lines are not included in the fit model. With proper inclusion of such
effects, residual uncertainty in the wavelength calibration is at the
1~cm/s level. Thus we conclude that while astro-comb line frequency
shifts caused by mirror dispersion and resultant wavelength dependence
of FPC finesse and phase deviation can be absorbed into the wavelength
calibration if reproducibility is all that is desired, at the few cm/s
level these corrections must be applied to the spectrograph
calibration to achieve accurate stellar radial velocity measurements.

\section{Conclusion}

In summary, we have demonstrated an \emph{in-situ} method to determine systematic shifts of astro-comb spectral lines due to imperfect suppression of source comb lines by the Fabry-Per\'ot filter cavity (FPC). This method involves measurement of FPC finesse as well as the phase deviation of all astro-comb lines over a bandwidth of 1000~\AA. Such measurements can be performed at a telescope with either an astrophysical spectrograph or a commercial optical spectrum analyzer of lower resolution. From the measured phase deviation, the dispersion of the FPC is derived and found to be consistent with that calculated from the intra-cavity air dispersion and the group delay dispersion of the FPC mirrors. Applying resultant corrections for shifts in the centroid of astro-comb spectral lines allows us to model the astro-comb spectrum as measured on an astrophysical spectrograph with accuracy to better than
0.1~MHz, i.e., 10 cm/s for measurement of a stellar radial
velocity. Such high accuracy is important for many applications of
astro-comb wavelength calibrators, such as the search for habitable exoplanets,
direct measurement of the expansion of the universe, and searches for a temporal variation of physical constants.



\begin{thebibliography}{10}
\newcommand{\enquote}[1]{``#1''}

\bibitem{Lovis2006}
C.~Lovis, M.~Mayor, F.~Pepe, Y.~Alibert, W.~Benz, F.~Bouchy, A.~C.~M. Correia,
  J.~Laskar, C.~Mordasini, D.~Queloz, N.~C. Santos, S.~Udry, J.-L. Bertaux, and
  J.-P. Sivan, \enquote{An extrasolar planetary system with three neptune-mass
  planets,} Nature \textbf{441}, 305--309 (2006).

\bibitem{Sandage1962}
A.~Sandage, \enquote{The change of redshift and apparent luminosity of galaxies
  due to the deceleration of selected expanding universes,} Astrophys. J.
  \textbf{136}, 319--333 (1962).

\bibitem{Loeb1998}
A.~Loeb, \enquote{Direct measurement of cosmological parameters from the cosmic
  deceleration of selected expanding universes,} Astrophys. J. \textbf{499},
  L111--L114 (1998).

\bibitem{Udem2002}
T.~Udem, R.~Holzwarth, and T.~W. Hansch, \enquote{Optical frequency metrology,} Nature \textbf{416}, 233--237 (2002).

\bibitem{Murphy2007}
M. T. Murphy, T. Udem, R. Holzwarth, A. Sizmann, L. Pasquini, C. Araujo-Hauck, H. Dekker, S. D'Odorico, M. Fischer, T. W. H\"ansch, and A. Manescau,
\enquote{High-precision wavelength calibration of astronomical spectrographs with laser frequency combs,} Mon. Not. R. Astron. Soc. \textbf{380}, 839-847 (2007).

\bibitem{Schmidt2007}
P.~O.~Schmidt, S.~Kimeswenger, and H.~U.~Kaeufl,
\enquote{A new generation of spectrometer calibration techniques based on optical frequency combs,}
in Proc. 2007 ESO Instrument Calibration Workshop (ESO Astrophysics Symposia series, Springer, in the press)

\bibitem{Araujo2007}
C.~Araujo-Hauck, L. Pasquini, A. Manescau, T. Udem, T. W. H\"ansch, R. Holzwarth, A. Sizmann, H. Dekker, S. D'Odorico, and M. T. Murphy,
\enquote{Future wavelength calibration standards at ESO: the laser frequency comb,} ESO Messenger \textbf{129}, 24--26 (2007).

\bibitem{Osterman2007}
S.~Osterman, S.~Diddmas, M.~Beasley, C.~Froning, L.~Hollberg, P.~MacQueen, V.~Mbele, and A.~Weiner,
\enquote{proposed laser frequency comb-based wavelength reference for high-resolution spectroscopy,} in Proc. SPIE, 6693, G1 (2007).

\bibitem{Li2008}
C.-H. Li, A.~Benedick, P.~Fendel, A.~Glenday, F.~K{\"a}rtner, D.~Phillips,
  D.~Sasselov, A.~Szentgyorgyi, and R.~Walsworth, \enquote{A laser frequency
  comb that enables radial velocity measurements with a precision of 1~cm~{s${}^{-1}$},} Nature \textbf{208}, 610--612 (2008).

\bibitem{Steinmetz2008}
T.~Steinmetz, T.~Wilken, C.~Araujo-Hauck, R.~Holzwarth, T.~W. Hansch,
  L.~Pasquini, A.~Manescau, S.~D'Odorico, M.~T. Murphy, T.~Kentischer,
  W.~Schmidt, and T.~Udem, \enquote{Laser frequency combs for astronomical
  observations,} Science \textbf{321}, 1335--1337 (2008).

\bibitem{Braje2008}
D. A. Braje, M. S. Kirchner, S. Osterman, T. Fortier, and S. A. Diddams,
\enquote{Astronomical spectrograph calibration with broad-
spectrum frequency combs,} Eur. Phys. J. D 48, 57 (2008).

\bibitem{Lovis2006b}
C.~Lovis, F.~Pepe, F.~Bouchy, G.~L. Curto, M.~Mayor, L.~Pasquini, D.~Queloz,
  G.~Rupprecht, S.~Udry, and S.~Zucker, \enquote{The exoplanet hunter HARPS:
  unequalled accuracy and perspectives toward 1~cm~{s${}^{-1}$} precision,}
  Proc. SPIE \textbf{6269}, 62690P1 -- 62690P23 (2006).


\bibitem{Udry2007}
S.~Udry, X.~Bonfils, X.~Delfosse, T.~Forveille, M.~Mayor, C.~Perrier,
  F.~Bouchy, C.~Lovis, F.~Pepe, D.~Queloz, and J.-L. Bertaux, \enquote{The
   HARPS search for southern extra-solar planets. xi. super-earths (5 and 8
  {M${}_\oplus$}) in a 3-planet system,} Astron. Astrophys. \textbf{469},
  L43--L47 (2007).

\bibitem{Ye05}
M.~J. Thorpe, R.~J. Jones, K.~D. Moll, J.~Ye, and R.~Lalezari, \enquote{Precise
  measurements of optical cavity dispersion and mirror coating properties via
  femtosecond combs,} Opt. Express \textbf{13}, 882--888 (2005).

\bibitem{Hansch06}
A.~Schliesser, C.~Gohle, T.~Udem, and T.~W. {H\"ansch}, \enquote{Complete
  characterization of a broadband high-ﬁnesse cavity using an optical
  frequency comb,} Opt. Express \textbf{14}, 5975--5983 (2006).

\bibitem{Furez2009}
G.~Furesz, \enquote{Design and application of high resolution and multiobject
  spectrographs: Dynamical studies of open clusters,} Ph.D. thesis, University
  of Szeged, Hungary (2008).

\bibitem{YarivPhotonics}
A.~Yariv and P.~Yeh, \emph{Photonics} (Oxford University Press, Oxford, 2006).

\bibitem{Ciddor1996}
P.~E. Ciddor, \enquote{Refractive index of air: new equations for the visible and near infrared,} Appl. Optics \textbf{35}, 1566--1573 (1996).

\bibitem{Kirchner2009}
M. S. Kirchner, D. A. Braje, T. M. Fortier, A. M. Weiner, L. Hollberg, and S. A. Diddams,
\enquote{Generation of 20 GHz, sub-40 fs pulses at 960 nm via repetition-rate multiplication,} Opt. Lett. \textbf{34}, 872 (2009).

\bibitem{Steinmetz2009}
T. Steinmetz, T. Wilken, C. Araujo-Hauck, R. Holzwarth, T. W. H\"ansch, and T. Udem,
\enquote{Fabry-Per\'ot filter cavities for wide-spaced frequency combs with large spectral bandwidth,} Appl. Phys. B \textbf{96}, 251 (2009)


\end{thebibliography}
\end{document}